\documentclass[aps,prl,showpacs,showkeywords,twocolumn,superscriptaddress]{revtex4}
\usepackage{bm,color,amsmath,amssymb,mathrsfs,latexsym,graphicx,psfrag}

\newcommand{\ket}[1]{|#1\rangle}
\newcommand{\ev}[1]{\langle #1 \rangle}
\newcommand{\pB}[0]{\mathbf p_{\mbox{\tiny B}}}
\newcommand{\bk}[0]{\mathbf k}
\newcommand{\bp}[0]{\mathbf p}
\newcommand{\bi}[0]{{\mathbf i}}
\newcommand{\bj}[0]{{\mathbf j}}
\newcommand{\bra}[1]{\langle #1|}


\begin{document}
\title{Detecting the Amplitude Mode of Strongly Interacting Lattice Bosons\\ by Bragg Scattering}

\author{Ulf Bissbort}
\affiliation{Institut f\"ur Theoretische Physik, Johann Wolfgang Goethe-Universit\"at, 60438 Frankfurt/Main, Germany}
\author{S\"oren G\"otze}
\affiliation{Institut f\"ur Laser-Physik, Universit\"at Hamburg, 22761 Hamburg, Germany}
\author{Yongqiang Li}
\affiliation{Institut f\"ur Theoretische Physik, Johann Wolfgang Goethe-Universit\"at, 60438 Frankfurt/Main, Germany}
\affiliation{Department of Physics, National University of Defense Technology, Changsha  410073, P. R. China}
\author{Jannes Heinze}
\affiliation{Institut f\"ur Laser-Physik, Universit\"at Hamburg, 22761 Hamburg, Germany}
\author{Jasper S. Krauser}
\affiliation{Institut f\"ur Laser-Physik, Universit\"at Hamburg, 22761 Hamburg, Germany}
\author{Malte Weinberg}
\affiliation{Institut f\"ur Laser-Physik, Universit\"at Hamburg, 22761 Hamburg, Germany}
\author{Christoph Becker}
\affiliation{Institut f\"ur Laser-Physik, Universit\"at Hamburg, 22761 Hamburg, Germany}
\author{Klaus Sengstock}
\affiliation{Institut f\"ur Laser-Physik, Universit\"at Hamburg, 22761 Hamburg, Germany}
\author{Walter Hofstetter}
\affiliation{Institut f\"ur Theoretische Physik, Johann Wolfgang Goethe-Universit\"at, 60438 Frankfurt/Main, Germany}

\begin{abstract}
We report the first detection of the Higgs-type amplitude mode using Bragg spectroscopy in a strongly interacting condensate of ultracold atoms in an optical lattice. By the comparison of our experimental data with a spatially resolved, time-dependent bosonic Gutzwiller calculation, we obtain good quantitative agreement. This allows for a clear identification of the amplitude mode, showing that it can be detected with full momentum resolution by going beyond the linear response regime. A systematic shift of the sound and amplitude modes' resonance frequencies due to the finite Bragg beam intensity is observed.
\end{abstract}

\maketitle

In recent years, remarkable progress has been made in the field of ultracold atoms, enabling the simulation of strongly interacting quantum systems beyond the scope of traditional solid state counterparts \cite{jaksch98, Bloch2008, greiner02}. In solid state systems, spectroscopic techniques such as ARPES and neutron scattering have been established as reference methods for providing energy and momentum resolved insight into the excitational structure of materials. Recently, spectroscopic techniques have also been applied successfully to ultracold atoms, such as RF spectroscopy \cite{Stewart2008}, lattice shaking \cite{Stoeferle2004, Schori2004}, as well as several experiments using Bragg spectroscopy \cite{Ernst09, Clement_Bragg,Fabbri2009, Stenger_Bragg, Papp_Bragg,Holland09, stamper_kurn99}. Initially, the latter was performed on weakly interacting condensates \cite{Stenger_Bragg, stamper_kurn99}, then extended to strong interactions without a lattice \cite{Papp_Bragg}. In more recent experiments, these studies have also been extended to ultracold atoms in optical lattices \cite{Ernst09,Clement_Bragg,Fabbri2009,Roth2004, van_Oosten2005, Pupillo2006, Rey2005, Krutitsky2010, Ye2010, Huber_PRB}, which opens up the possibility of studying a number of models with strong correlations from condensed matter theory. Up to now, these Bragg spectroscopic experiments have, however, been focused on weakly interacting condensates \cite{Ernst09,Clement_Bragg,Fabbri2009} or the Mott insulating (MI) \cite{Clement_Bragg} regime. In this letter, we investigate the excitational structure of a strongly interacting lattice superfluid (SF) and, for the first time, clearly identify the recently described \emph{amplitude mode} \cite{Huber2008,Huber_PRB,Huber_PhD,gapped_modes,Menotti08_Ohashi2006}.

At sufficiently high lattice depth, the atoms are well described by the Bose-Hubbard model (BHM)
\begin{equation*}
\vspace{-2.2mm}
	\mathcal{H}=-J \sum_{\ev{i,j}} ( b_i^\dag b_j^{\phantom{\dag}} +  \mbox{h.c.})+\sum_i (\epsilon_i-\mu) {b}_i^\dag b_i^{\phantom{\dag}} + \frac{U}{2}\sum_{i} b_i^\dag b_i^\dag b_i^{\phantom{\dag}} \! b_i^{\phantom{\dag}},	
	\vspace{-1mm}
\end{equation*}
where ${b}_i^\dag$ creates an atom at lattice site $i$, the tunneling between nearest neighboring sites is characterized by a tunneling matrix element $J$ and the on-site energy shift an atom experiences at a given site in the presence of $n$ other atoms is given by $Un$, with the interaction parameter $U$. A local energy offset is accounted for by $\epsilon_i$ and $\mu$ denotes the chemical potential. The crucial parameter for realizing different regimes is the ratio $U/J$. In the Bogoliubov regime $U/J\ll 1$, the gapless sound mode, corresponding to the excitation of Bogoliubov quasiparticles in a lattice has been investigated experimentally \cite{Ernst09,Clement_Bragg,Fabbri2009}. Intermediate lattice depths allow for the realization of a strongly interacting SF beyond the realm of Bogoliubov theory, exhibiting a rich excitational structure: In addition to the gapless sound mode, the existence of the gapped `amplitude' mode in the BHM (within the lowest band), generated by a physically similar mechanism as the Higgs boson in high energy physics \cite{Huber_PRB,Huber2008,Huber_PhD}, has  been a topic of high interest in recent literature \cite{Huber2008,Huber_PRB,Huber_PhD,gapped_modes,Menotti08_Ohashi2006}. However, linear response calculations in the perturbative limit have suggested that this mode cannot be addressed in a momentum-resolved fashion with Bragg spectroscopy \cite{Huber2008} and there has been no clear experimental signature in previous measurements \cite{Clement_Bragg}. To bridge the gap between existing idealized theory predictions and our experimental observations, we address a number of important experimental effects in our simulations: 1) the high probing beam intensity; 2) spatial inhomogeneities, such as the harmonic trapping potential breaking the translational symmetry and leading to a broadening in $k$-space; 3) strong interactions in the SF requiring a treatment beyond Bogoliubov theory; 4) the short probing pulse time leading to a broadened signal in $\omega$-space. Each of these effects can modify the resulting measurement and a comprehensive analysis has  not been performed to date.

In our experiment, $^{87}$Rb atoms are cooled in a shallow magnetic trap with $\omega=2\pi\cdot(16,16,11)\,$Hz, forming a Bose-Einstein condensate before a 3D cubic optical lattice with a spacing of $a=515$nm is slowly ramped up to a final intensity of $s$ recoil energies $E_r$, as described in Ref.~\cite{Ernst09}. This transfers the atoms into a condensed state in the lowest band of the lattice, where the system is well described by the BHM and the $s$-wave interaction through the background scattering length is parametrized by the interaction constant $U$. Subsequently, two Bragg laser beams with a slight frequency detuning $\omega_B$ but essentially the same wavelength $\lambda=781.37$nm (i.e. $|\omega_B|\ll c/\lambda$), lying in the $x$-$y$-plane of the optical lattice at a coincident angle $\theta_B=45^\circ$, are applied. This allows the atoms to undergo a two-photon process, in which the momentum kick an atom experiences is given by $|\pB|=(4 \pi /\lambda) \sin(2 \theta_B)$. Our specific experimental setup allows the system to be probed along the nodal direction. For brevity, all dispersion relations and results shown in this paper are along this line, connecting the $\Gamma=(0,0,0)$ and $M=(1,1,0)$ points in the first Brillouin zone (BZ).

\begin{figure}[t!]
\begin{center}
\includegraphics[width=\linewidth]{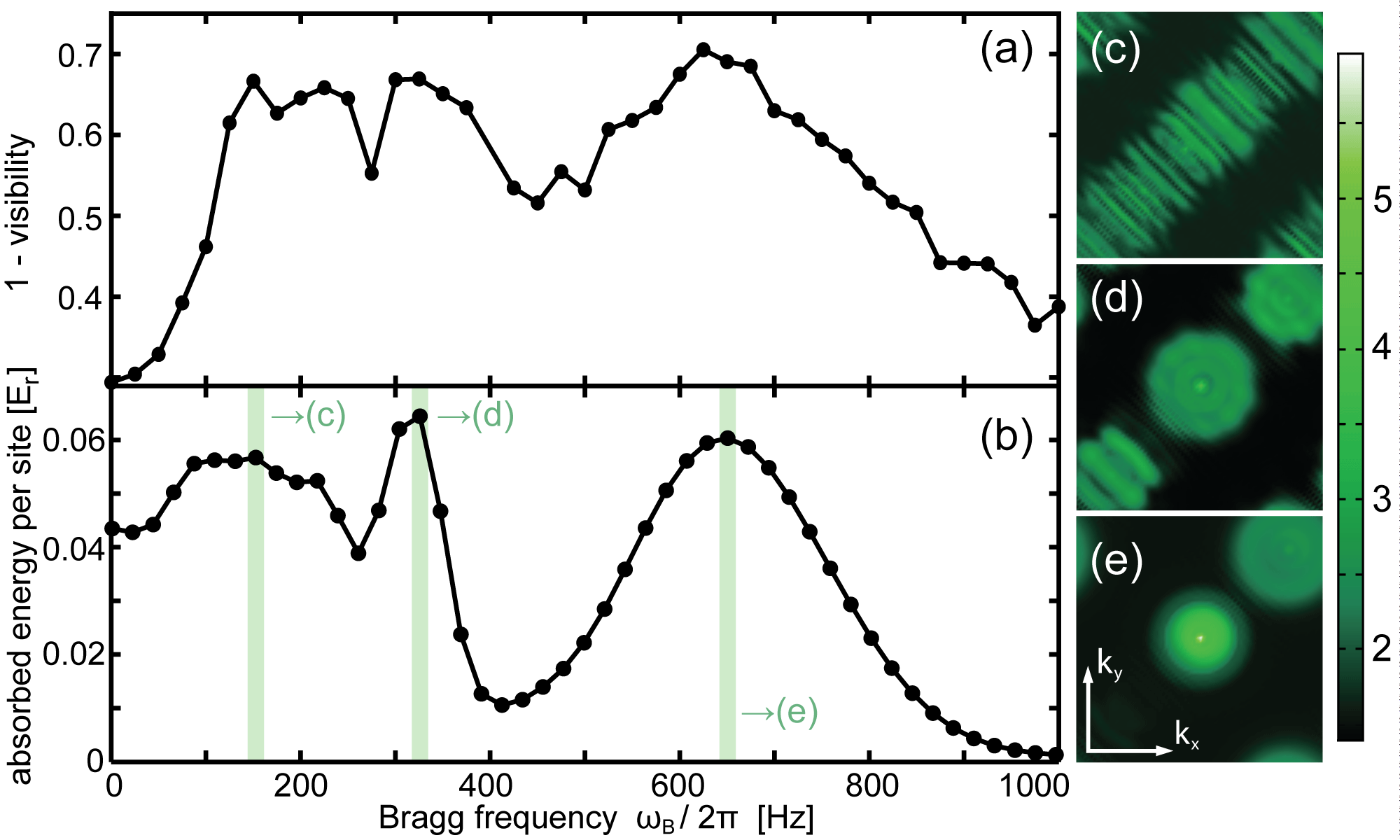}
\vspace{-4mm}
\caption
 {\label{fig:experiment} 
 (Color online) Comparison of the experimental visibility (a) and theoretically predicted energy absorption (b) at $|\pB|= \pi/a$ using a Blackman-Harris pulse of $10$ms in an optical lattice with $s=13$ in a 3D trap. A maximum intensity $V=0.27E_r$ and the experimentally determined total particle number $N_{\mbox{\tiny tot}}=5\cdot 10^{4} \; \pm33\%$ and $\omega=2\pi\cdot(26,26,21)\,$Hz trapping frequency were used for (b), leading to a maximum central density $n=1.05$. The lower peak is the sound mode, mainly broadened by the high intensity of the Bragg beam to lower frequencies. The upper peak at $~650$Hz is the gapped amplitude mode, broadened mainly by the trap. Figures (c,d,e) show the theoretically predicted trap broadened logarithmic quasi-momentum distributions in the first Brillouin zone at the frequencies marked by the green lines in (b).}
\vspace{-9mm}
\end{center}
\end{figure}

Within a classical treatment of the laser field and using the correspondence principle, the effect of the time-dependent Bragg field on the atoms is theoretically described by the single particle operator $\mathcal B(t)=\frac V 2 \left( e^{-i\omega_B t}\rho_{\pB}^\dag+e^{i\omega_B t}\rho_{\pB}^{\phantom{\dag}} \right)$, corresponding to a propagating sinusoidal potential with wave vector $|\pB|$, where $V$ denotes the Bragg intensity and we use units of $\hbar=1$. In free space the operator  $\rho_{\pB}^\dag=\sum_\bp a_{\bp+\pB}^\dag a_\bp^{\phantom{\dag}}$ acts as a translation operator in momentum space and simply transfers atoms into higher momentum states $\pB$, if energetically allowed, where $a_\bp$ is the annihilation operator for a momentum state $\bp$. However, in the presence of an optical lattice interactions are intensified and the multi-band structure and periodicity of the BZ invalidate this intuitive picture: multiple scattering events are enhanced and may lead to the occupation of a broad distribution of momentum components [Fig.~\ref{fig:experiment}(c)-(d)]. Moreover, strong interactions require an analysis in terms of a renormalized quasi-particle picture. In this Letter we focus on the physics within the lowest band, requiring all relevant energy scales to be lower than the band gap.

To incorporate the Bragg operator into our dynamic Gutzwiller (GW) calculation, it is transformed into Wannier space via the unitary transformation obtained from a band structure calculation as explicated in \cite{supplementary_material}, Appendix A. This leads to a lowest band representation $\rho_{\pB}^\dag=\sum_{ {\mathbf i}, {\mathbf j}}\rho_{{\mathbf i},{\mathbf j}} b_{{\mathbf i}}^{{\dag}} b_{{\mathbf j}}^{\phantom{\dag}}$, with the exact intra-band matrix elements $\rho_{{\mathbf i},{\mathbf j}}$ treated beyond the on-site and nearest neighbor approximation (decaying exponentially with $|{\mathbf i} - {\mathbf j}|$), where ${\mathbf i},{\mathbf j}$ denote the site indices. Within bosonic Gutzwiller theory, the variational ansatz for the many-body state consists of a single tensor product of states at each site $\ket{\psi(t)}=\prod_{\otimes {\mathbf i}}\ket{\phi_{\mathbf i}(t)}_{\mathbf i}$, which correctly recovers both the atomic limit $U/J\to \infty$ and time-dependent Gross-Pitaevskii theory within a coherent state description for weak interactions and it becomes exact in high spatial dimensions. For a strongly interacting condensate in the vicinity of the Mott transition, it furthermore includes the physics of the effective theories by Huber et al. \cite{Huber2008,Huber_PRB}. For a given trap geometry and experimental parameters, the ground state is determined and subsequently time evolved in the presence of the Bragg beam. The equations of motion are determined by minimizing the action (including the time-dependent Bragg operator) and are equivalent to the time evolution generated by a set of effective local Hamiltonians, coupled non-linearly to the states at other sites (see \cite{supplementary_material}, Appendix B). 

The pulse shape investigated theoretically here, is the square pulse where the Bragg intensity $V$ is constant over a fixed time interval $t$. This leads to the characteristic sinc$^2$ response in frequency space, as is to be expected from time-dependent perturbation theory and can be seen in Fig.~\ref{fig:absorption_spectra}. To minimize the oscillatory response for a restricted Bragg pulse time, a Blackman-Harris pulse \cite{Harris78} is used to obtain the central results shown in Fig.~\ref{fig:experiment}, both in experiment and theory. Since the energy absorbed from the Bragg pulse leads to a depletion of the condensate after rethermalization, the former is monotonically related to the visibility \cite{Bloch2008} and it is useful to compare these two quantities.  The lower peak at $\sim\!200\mbox{Hz}$ in the spectra is the trap- and intensity-broadened sound mode, whereas the higher peak at $\sim\!650\mbox{Hz}$ is the amplitude mode, broadened mainly by the strong density dependence. 

\begin{figure}[t!]
\begin{center}
\includegraphics[width=\linewidth]{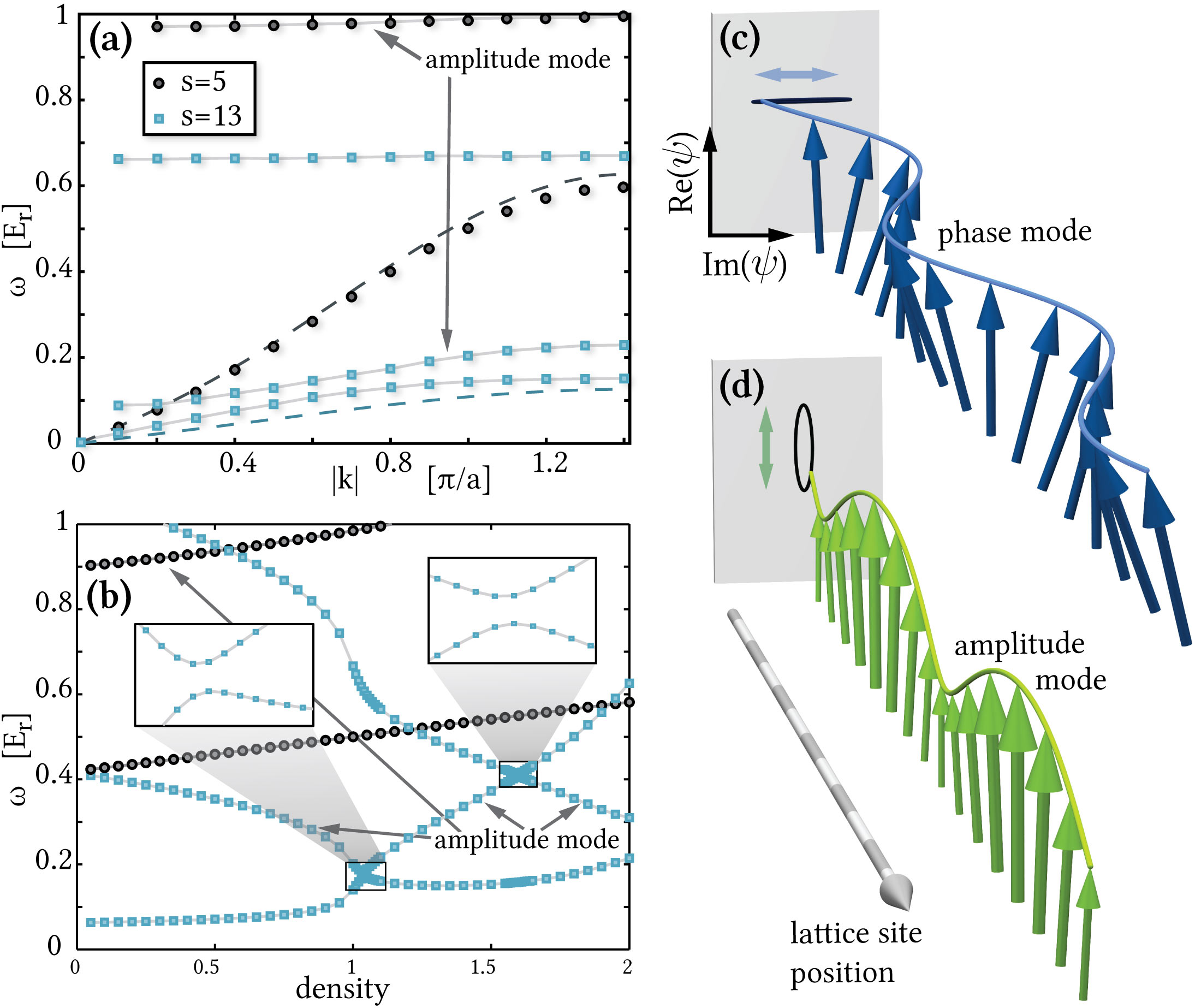}
\vspace{-4mm}
\caption
 {\label{fig:disp_rel}
   (Color online) (a,b): Theoretical dispersion relations for the homogeneous system in the linear response limit at $n=1$ (a) and density-dependence at $|\bk|=\pi/a$ (b) in the SF for weak ($s=5$, black circles) and strong ($s=13$, blue squares) interactions. Corresponding Bogoliubov results \cite{van_Oosten_2001} are shown as black dashed lines in (a). (c,d): Illustration of the order parameter for a coherent excitation of the phase (sound) mode (c) and the amplitude mode (d) in a homogeneous condensate at $\bk=(0.8/a,0,0)$ and $s=13$. The projection of all $\psi_{\mathbf i}$'s in the complex plane is shown by the black ellipses: for the sound mode (c), the oscillation is almost exclusively in the tangential, for the amplitude mode (d) mainly in the radial (i.e. in the amplitude) direction.}
\vspace{-9mm}
\end{center}
\end{figure}

\begin{figure*}[t!]
\begin{center}
\includegraphics[width=\linewidth]{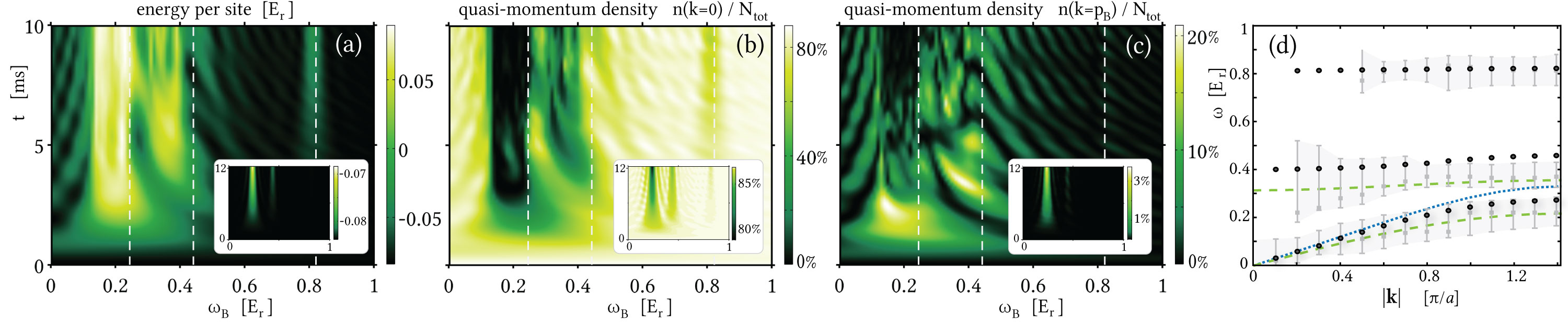}
\vspace{-7mm}
\caption
{\label{fig:absorption_spectra}
(Color online) Energy absorption (a) and quasi-momentum density (b),(c) spectra for a square pulse at high intensity $V=0.1E_r$ (insets: weak intensity $V=0.005 E_r$) in the intermediate interaction $s=9$ regime and for Bragg momentum $|\pB|=\pi/a$. The resonance frequencies predicted from the maxima of the high intensity energy absorption spectra (plotted as gray squares in (d)) contain a systematic uncertainty quantified by the FWHM of the pulse after $10\mbox{ms}\approx3.26/J$ indicated by the error bars and shaded region in (d). The comparison with the true quasiparticle energies (dashed white lines in (a-c), black circles in (d)) reveals significant discrepancies. For comparison in (d): The blue dotted line is the Bogoliubov result, the green dashed lines are the results from Ref.~\cite{Huber2008} for the amplitude and sound modes ($\omega(\bk)\!=\!\psi_0 \sqrt{2Un \epsilon_\bk}$ with $\psi_0$ determined by GW).
\vspace{-9mm}
}
\end{center}
\end{figure*}

While exposed to the Bragg lasers, atoms are continuously transferred between different quasi-momentum states, with $\bk\!=\!0\!\to\!\pB$ initially being the dominant transition at weak interactions. With increasing $U/J$, backscattering transitions are enhanced and at longer times higher order transitions also become relevant. This can also be seen from the physical momentum distribution $n(\bp)=\langle a_\bp^\dag a_\bp^{\phantom{\dag}} \rangle$, which is directly related to the quasi-momentum distribution, as shown in Fig.~\ref{fig:experiment}(c)-(e). In the low intensity and long-time limit, Bragg spectroscopy directly probes the dynamic structure factor. For fixed $\pB$, the various quasiparticle energies can be determined from the strongest loss in the momentum component $n(\bk=0)$, gain in $n(\bk=\pB)$, energy absorption or reduction in the condensate fraction as a function of the frequency $\omega_B$, as shown in Fig.~\ref{fig:experiment}(a),(b). At large $s$, an additional complication arises in experiments: since the condensate is strongly depleted, the time of flight images are very similar to those of a thermal cloud within the signal to noise ratio. Thus, the lattice depth is ramped down linearly over $10$ms to $s\!=\!10$ after  exposure to the Bragg beams. Subsequently, the visibility, shown in Fig.~\ref{fig:experiment}(a), is extracted from the time of flight image of the equilibrated atoms and is monotonically related to the absorbed energy. Determining these resonance positions for a range of different momenta $\pB$ leads to the dispersion relations with the Bogoliubov, amplitude and higher gapped modes shown in Fig.~\ref{fig:disp_rel} and compared with other theoretical results. A probing beam at resonance with a collective mode frequency induces time- and position-dependent oscillations of the density and the spatial order parameters $\psi_{\mathbf i}=\langle b_{\mathbf i} \rangle$. In a theoretical description, these excitations correspond to coherent states of the respective quasiparticle, i.e. the most classical excitation, and are graphically illustrated in Fig.~\ref{fig:disp_rel}(c),(d): a coherent Bogoliubov excitation leads to a dominant spatial and temporal oscillation of the phase of $\psi_{\mathbf i}$ (which becomes pure for $k\!\to\!0$) and a density wave, whereas an excitation of the amplitude mode leads mainly to an oscillation of the amplitude of $\psi_{\mathbf i}$ and the density modulation is strongly suppressed. The oscillation of $|\psi_{\mathbf i}|$ at constant density can thus be understood as a local periodic transfer of particles between the condensate and the non-condensate. 
In contrast to the weakly interacting case, where the quasiparticle energies of the different modes depend approximately linearly on the density (black dotted lines in Fig.~\ref{fig:disp_rel}(b)), the dependence in the strongly interacting case is highly non-trivial. The strong dependence can be understood from the excitational particle and hole branches, which may cross each other in the Mott insulator: crossing the phase transition into the SF, the emerging condensate couples the particle/hole branches in the equations of motion, hybridizing these and leading to avoided mode crossings at the previous intersection points, as is shown by the blue squares in Fig.~\ref{fig:disp_rel}(b). For all $\bk$, $U/J$ and densities in the SF, the sound (amplitude) mode remains the energetically lowest (second lowest) lying mode. Comparing our theoretical results with Bogoliubov theory \cite{van_Oosten_2001} (black dashed line in Fig.~\ref{fig:disp_rel}(a)), excellent agreement is obtained in the weakly interacting limit. At intermediate interactions $s=9$ ($U/J\approx8.55$) and density $n=1$ shown in Fig.~\ref{fig:absorption_spectra}(d)), neither Bogoliubov theory (dotted blue line), nor the theory presented by Huber et al. \cite{Huber2008} for strong interactions (dashed green lines) apply and deviate from our results. Here, the dispersion relation obtained by the dynamic GW method (black circles in Fig.~\ref{fig:absorption_spectra}(d)) remains valid and continuously connects these two limiting theories. 

An essential effect that has to be considered in a realistic modeling of Bragg spectroscopy is the finite intensity of the probing beam. This is particularly important for strong optical lattices, where the typical time scale $1/J$ grows exponentially with $s$. As the pulse time is restricted by decoherence, a increasing intensity $V$ is required for strong lattices. The analysis of this effect requires a treatment beyond the linear response of the system (i.e. not contained in the dynamic structure factor), as shown in the spectra of the full time-dependent GW calculation in Fig.~\ref{fig:absorption_spectra}. Whereas the response in the limit of very small $V$ shown in the insets of Fig.~\ref{fig:absorption_spectra}(a)-(c) is given by $\delta$-shaped peaks as expected, there is a drastic non-trivial broadening of the different peaks for typical experimental intensities $V\approx 0.1 E_r$, shown in the respective main figures. This indicates a breakdown of the non-interacting quasiparticle picture of the BHM due to the large $V$. Whereas the amplitude mode's signature is generally stronger in the energy- and $n(\bk=0)$ than in the $n(\bk=\pB)$ profile, the scaling of its spectral weight is nonlinear in V (and thus beyond linear response in this large $V\cdot t$ regime), as shown in comparison of Figs.~\ref{fig:absorption_spectra}(a),(b) and the respective insets. 

At high intensity $V$ the spectra are not only broadened, but the supposed resonance frequencies of all modes (gray squares in Fig.~\ref{fig:absorption_spectra}(d)) are systematically shifted to lower frequencies with respect to the true quasiparticle energies (indicated by dashed white lines in Fig.~\ref{fig:absorption_spectra}(a)-(c) and circles in d)), consistent with RPA \cite{Menotti08_Ohashi2006}), i.e. the quasiparticle energies are renormalized by the interaction induced by $V$. The error bars and shaded areas in Fig.~\ref{fig:absorption_spectra}(d) indicate the FWHM of the energy absorption profile after $t=10\mbox{ms}$, quantifying the systematic uncertainty in the extracted energies. To the best of our knowledge, this has not been considered in the analysis of experimental data thus far. Two further effects accounted for in our calculation are the frequency broadening due to the finite pulse time, as well as the inhomogeneous trapping potential. A shallow trap and low filling $n\lesssim 1.05$, due to the strong density dependence of the mode frequencies, are crucial for an unambiguous identification of the amplitude mode. We stress that only by taking all these effects into account, the good quantitative agreement, shown in Fig.~\ref{fig:experiment}(b), between theory and experiment in the spectra is achieved. In \cite{supplementary_material}, we point out the underlying connection to lattice amplitude modulation, and furthermore perform a time-dependent calculation for the experiments \cite{Stoeferle2004,Schori2004} in 3D, finding good agreement in the absorption peak frequency at different $s$.

In conclusion, we have experimentally observed the gapped amplitude mode of the BHM in the strongly interacting superfluid regime using Bragg spectroscopy. Good quantitative agreement between the experimental visibility and the theoretically predicted energy absorption from a time-dependent bosonic Gutzwiller calculation is found, but only when taking the full spatial trap profile, finite pulse time and high intensity of the probing beam into account. This shows that Bragg spectroscopy is a suitable method for probing not only the quasiparticle structure of Bogoliubov mode with full momentum resolution, but also of the more exotic collective amplitude mode excitation. For a clear signal of the latter in a strongly interacting SF, a shallow trap on the experimental side and a theoretical treatment beyond the perturbative linear response regime are essential. Whereas a finite Bragg beam intensity is vital for a clear spectroscopic response of the amplitude mode, it leads to a renormalization of the sound- and amplitude mode resonance energies, which  has to be accounted for in a quantitative comparison of experiment and theory.

\begin{acknowledgments}
We thank P.~T.~Ernst and A.~Pelster for stimulating discussions. This work was supported by the German Science Foundation (DFG) via Forschergruppe FOR 801. Y.~L. was supported by the China Scholarship Fund. Calculations were performed at the CSC Frankfurt.
\end{acknowledgments}

\newpage
\phantom{.}
\newpage

\appendix
\section{Appendix A: The Bragg Operator}
\label{Bragg_op}
In this section we derive the Wannier representation of the single particle operator associated with the Bragg beam. This operator $\mathcal B(t)$, which effectively describes the virtual absorption and stimulated emission of a photon via an intermediate level is most naturally expressed in physical momentum space, where it acts as a superposition of two translation operators $\rho_{\pB}^{\phantom{\dag}}$ and $\rho_{\pB}^{\dag}$ with time-dependent complex phases. For the simulation within the frame of the lowest band Bose-Hubbard model, we seek a representation of this operator, which goes beyond the lowest order description of a moving wave for the local on-site energies. Note that whereas the Bragg operator $\mathcal B(t)=\frac V 2 \left( e^{-i\omega_B t}\rho_{\pB}^\dag+e^{i\omega_B t}\rho_{\pB}^{\phantom{\dag}} \right)$ is not separable within the different spatial dimensions, the two constituents $\rho_{\pB}$ and $\rho_{\pB}^{{\dag}}$ are very well separable. We therefore first focus on expressing $\rho_{\pB}^\dag$ in the Wannier basis for the 1D case as $\rho_{\pB}^\dag=\sum_{\bi,\bj}\rho_{\bi,\bj} b_{\bi}^{{\dag}} b_{\bj}^{\phantom{\dag}}$, before composing the Bragg and lattice amplitude modulation operators for the full 3D case in terms of these matrix elements.

One starts with the single particle basis transformation between the Bloch and physical momentum states
\begin{equation}
	\ket {k,\alpha}=\sum_{n=-\infty}^\infty  c_n^{(\alpha,k)} \ket{p=2n q_l + k}
\end{equation}
where $q_l=\pi/a$ is the lattice momentum and $\alpha$ the band index. The coefficients $c_n^{(\alpha,k)}$ can be explicitly obtained by diagonalizing the lattice Hamiltonian in quasi-momentum space. 
Choosing the appropriate normalization, these coefficients constitute a unitary matrix for a parametrically fixed quasi-momentum $k$
\begin{align}
	\sum_n {c_n^{(\alpha,k)}}^* {c_n^{(\alpha',k)}}&=\delta_{\alpha,\alpha'}\\
	\sum_\alpha {c_n^{(\alpha,k)}}^* {c_m^{(\alpha,k)}}&=\delta_{n,m}.
\end{align}

Furthermore it is useful to define the functions $N(p)=\left[ \frac{p}{2 q_l}  \right]$ and $K(p)=p-2q_l N(p)$ for the transformation from physical momentum to Wannier space, which allows the transformation between Bloch and true momentum states to be expressed as
\begin{equation}
	{b_{K(p)}^{(\alpha)}}^\dag =\sum_{m=-\infty}^\infty {c_{N(p+2q_l m)}^{(\alpha,K(p))}} \, a_{p+2q_l m}^\dag.
\end{equation}
As required by symmetry, this expression is manifestly invariant under $p\mapsto p+2 m q_l $ with $m\in \mathbb{Z}$.

For the basis transformation and subsequent projection onto the lowest band (i.e. considering only terms with $\alpha=0$) of the lattice along the dimension $d$ with $L^{(d)}$ lattice sites, we define the function
\begin{equation}
M(|\bi-\bj|, q)= \frac{1}{L^{(d)}} \sum_{p}   e^{ia p \,|\bi-\bj|} \, c_{N(p+q)}^{{(\alpha=0, k=K(p+q) )}^*} c_{N(p)}^{{(\alpha=0, k=K(p) )}} ,
\end{equation}
which fulfills the relation $M(-|\bi-\bj|, q)=M^*(|\bi-\bj|, q)$. The full 3D operator can then conveniently be expressed in terms of this function as
\begin{align}
\rho_{\bi,\bj}^\dag =e^{i a \pB \cdot \bi}\prod_{d=1}^3 M((\bi-\bj)\cdot \mathbf{e}_d,\pB\cdot \mathbf{e}_d),
\end{align}

where $\bi,\bj$ are the 3D lattice site vectors, containing the integer site numbering of the  cubic lattice along each dimension and $\mathbf{e}_d$ is the unit vector along the $d$-th dimension. Note that these matrix elements do not only depend on the Bragg momentum $\pB$, but also implicitly on the lattice depth $s$.

A short calculation explicitly verifies the property $\rho_{\bi,\bj}(p_B=0)=\delta_{\bi,\bj}$ along one dimension, i.e. different layers are not coupled by the Bragg process if $\pB$ lies within a plane parallel to these layers.
To first order, the Bragg operator locally corresponds to a sinusoidal potential (shift of effective chemical potential) moving along the direction $\pB$, but higher orders also give rise to nearest neighbor and longer range hopping. Usually these higher order terms are neglected, but their significance increases in certain regimes, such as when $\pB$ approaches $2 q_l$ (see Appendix C).
In Fig.~1 the modulus of the matrix elements $\rho_{\bi,\bj}$ is plotted as a function of the discrete lattice distance $|\bi-\bj|$ for different lattice depths $s$ and Bragg momenta $p_B$.

\begin{figure}[h]
\begin{center}
\includegraphics[width=\linewidth]{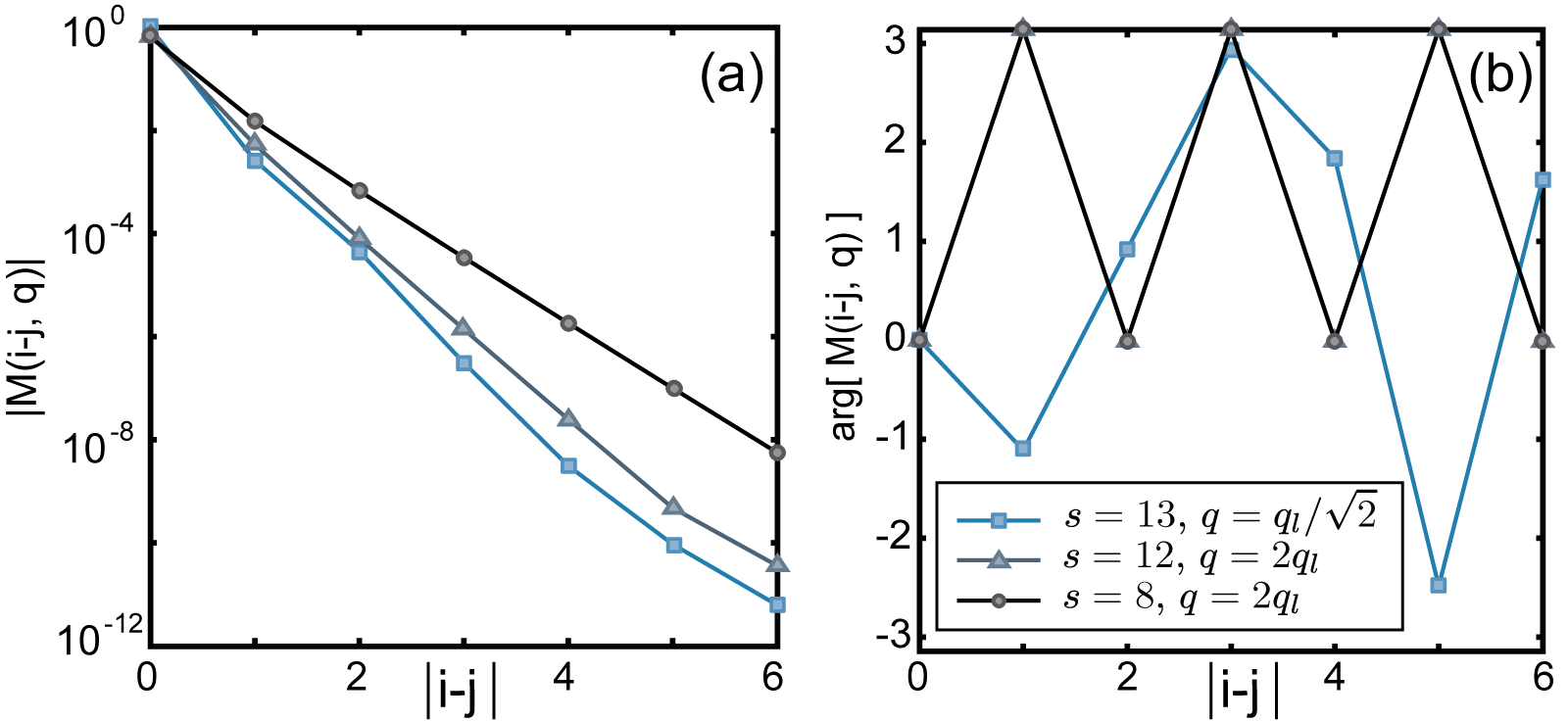}
\vspace{-4mm}
\caption
 {\label{fig:rho_me} 
 Dependence of complex matrix elements on the discrete lattice distance $|\bi-\bj|$ (subfigure a: modulus, subfigure b: arg). The values for $q=q_l/\sqrt{2}$ and $s=13$ are relevant for the Bragg experiment addressed in the main text, while the value of $q=2q_l$ is relevant for lattice modulation spectroscopy, discussed in Appendix C. The behavior of the complex angle of $M(\bi-\bj)$ is determined by $q$ and alternates in a regular fashion for $q=2q_l$ (here both curves lie on top of one another) in contrast to the case of $q=q_l/\sqrt{2}$. The long-distance exponential decay constant is primarily governed by the lattice depth $s$.
 }
\vspace{-9mm}
\end{center}
\end{figure}

\section{Appendix B. Time Evolution within the Dynamic Gutzwiller Approach}

The variational ansatz within the bosonic Gutzwiller approach for the state on the total lattice is the direct product state $\ket{\psi(t)}=\prod_{\otimes \bi} \ket{\phi_\bi(t)}_\bi$ with the most general form of a local lattice state. This can be expanded in any local basis, such as the local Wannier Fock basis $\ket{\phi_\bi(t)}_\bi=\sum_n f_n^{(\bi)}(t)\ket{n}_\bi$, where  $f_n^{(\bi)}(t)$ are now time-dependent fields within the dynamic approach. As the action
\begin{equation}
\mathcal{S}=\int dt \, \left[- i \bra{\psi(t)} (\partial_{t} \ket{\psi(t)})+  \bra{\psi(t)} \mathcal{H}_{\mbox{\tiny BH}} + \mathcal{B}(t)  \ket{\psi(t)}    \right]
\end{equation}
becomes extremal for the physical path and the energy function is a continuous function of the amplitudes $f_n^{(\bi)}(t)$, the variation with respect to any of the fields ${f_n^{(\bi)}}^*(t)$ leads to the equations of motion. The dynamics defined by these equations for $f_n^{(\bi)}(t)$ is fully equivalent to a physically more intuitive formulation, where the time evolution of the local state $\ket {\phi_\bi(t)}_\bi$ (at site $\bi$) is generated by the effective local Hamiltonian 
\begin{equation}
\label{Eq:eff_local_H}
\mathcal{H}_{\bi}=\frac U 2 b_\bi^\dag  b_\bi^\dag b_\bi \,b_\bi-[\mu-B_{\bi,\bi }(t)]\,  b_{\bi}^\dag  b_\bi -\eta^*(t)b_\bi -\eta(t)  b_\bi^\dag.
\end{equation}
Within this picture, all local on-site operators (or sums thereof) are accounted for exactly, whereas the non-local terms of any single particle operator are decoupled as
\begin{equation}
\sum_{\stackrel{\bi,\bj}{{\bi \neq \bj}} }A_{{\bi,\bj}} b_{\bi}^{\dag}b_{\bj}^{\phantom{\dag}} \mapsto \sum_{\stackrel{\bi,\bj}{{\bi\neq \bj}} }A_{{\bi,\bj}} ( \ev{ b_{\bi}^{\dag}}  b_{\bj}^{\phantom{\dag}}+  b_{\bi}^{\dag} \ev{ b_{\bj}^{\phantom{\dag}}} - \ev{ b_{\bi}^{\dag}}  \ev{ b_{\bj}^{\phantom{\dag}} } ).
\end{equation}
In our specific case this leads to the time-dependent complex parameters $\eta_{\bi}(t)=J\sum_{\bj \in n(\bi)} \psi_{\bj}-\sum_{\bj\neq \bi}B_{\bi,\bj}\,\psi_{\bj}$ in Eq.~(\ref{Eq:eff_local_H}), where $\psi_{\bi}(t)=\ev{b_{\bi}}$ is the local order parameter, directly related to the Wannier space representation of the condensate state and $n(\bi)$ denoting the set of all sites, which are nearest neighbors of sites of $\bi$. The parameter $\eta_{\bi}(t)$ thus accounts for both the non-linear nearest neighbor coupling through the hopping term in $\mathcal{H}_{\mbox{\tiny BH}}$, as well as the long range coupling to all other sites in the plane, mediated by $\mathcal{B}(t)$. 

By choosing sufficiently small time steps during the evolution, the numerical error is well controlled. Furthermore, various constants of motion (e.g. the total particle number, the energy for time independent Hamiltonians, the quasi-momentum for translationally invariant lattice potentials, etc.) are also conserved during the dynamic Gutzwiller time evolution. Numerically, realistically large (i.e. of the order $~10^6$ sites), strongly interacting non-equilibrium 3D lattice systems can be treated for all experimentally relevant time scales within this approach.

At every point in time physical quantities, such as the total energy (evaluated in the pure BHM Hamiltonian) $E(t)=\bra{\psi(t)} \mathcal{H}_{\mbox{\tiny BH}} \ket{\psi(t)} $, the condensate fraction or the quasi-momentum distribution $n(\mathbf{k},t)=\bra{\psi(t)} b_k^\dag b_{k}^{\phantom{\dag}} \ket{\psi(t)}$  are directly accessible from the many-particle state. The latter is easily related to the physical momentum distribution by $	\ev{a_p^\dag a_p^{\phantom{\dag}}}=\sum_{\alpha, \alpha'} {c_{N(p)}^{(\alpha, K(p))}}^* {c_{N(p)}^{(\alpha', K(p))}} \ev{  {b_{K(p)}^{(\alpha)}}^{\dag}  {b_{K(p)}^{(\alpha')}}^{\phantom{\dag}} }$ and consists of scaled replicas in the higher Brillouin zones within the lowest band approximation.

\section{Appendix C. Lattice Modulation Spectroscopy}
\label{sec_lat_mod}
To draw a connection to the lattice modulation experiments by St\"oferle et al. [5,6], we furthermore performed a time-dependent Gutzwiller calculation based on their experimental parameters. Usually, lattice modulation is understood as a periodic modulation of the nearest neighbor hopping element $J$, whereas Bragg spectroscopy is thought of as a spatial sinusoidal wave in the on-site energies, moving through the lattice at a velocity $\frac{\omega_B}{p_B}$. Both of these approximations are only true up to lowest order. Here we shortly derive the exact description of these two processes (as used in our numerical implementation) and also point out a fundamental connection between lattice amplitude modulation and Bragg spectroscopy.

It is useful to express the optical lattice potential operator in terms of the momentum shift operator $\rho_p$ (up to a constant) as $V_{\mbox{\tiny lat}}=\frac{s E_r}{4} (\rho_{2q_l}^\dag+\rho_{-2q_l}^\dag)$. During lattice modulation spectroscopy, the lattice amplitude is temporally modulated with a frequency $\omega_m$ and the perturbing operator becomes
\begin{equation}
	\hat V_{\mbox{\tiny mod}}=\frac{V_m}{4} \, \cos(\omega_m t) \, (\rho_{2q_l}^\dag+\rho_{-2q_l}^\dag).
\end{equation}

Decomposing the cosine into its two complex exponential constituents, this form of the lattice amplitude modulation operator reveals that it corresponds exactly to the sum of two Bragg processes with opposite frequencies $\omega_B=\pm \omega_m$ at fixed momenta $p= \pm 2 q_l$ in the direction of the modulated lattice laser beam. However, in contrast to Bragg spectroscopy at $q\neq 2mq_l$ with $m\in \mathbb Z$, where the most relevant lowest order term is the offset of the local potential, this term is irrelevant for lattice amplitude modulation (at $q=2 q_l$), where this potential offset is equal for every site and thus only leads to an irrelevant global phase. Hence, the physically relevant lowest order term for lattice amplitude modulation is the temporally modulated coupling to the nearest neighboring sites, i.e. a modulation of $J$.

\begin{widetext}
\begin{figure}[ht]
\includegraphics[width=17cm]{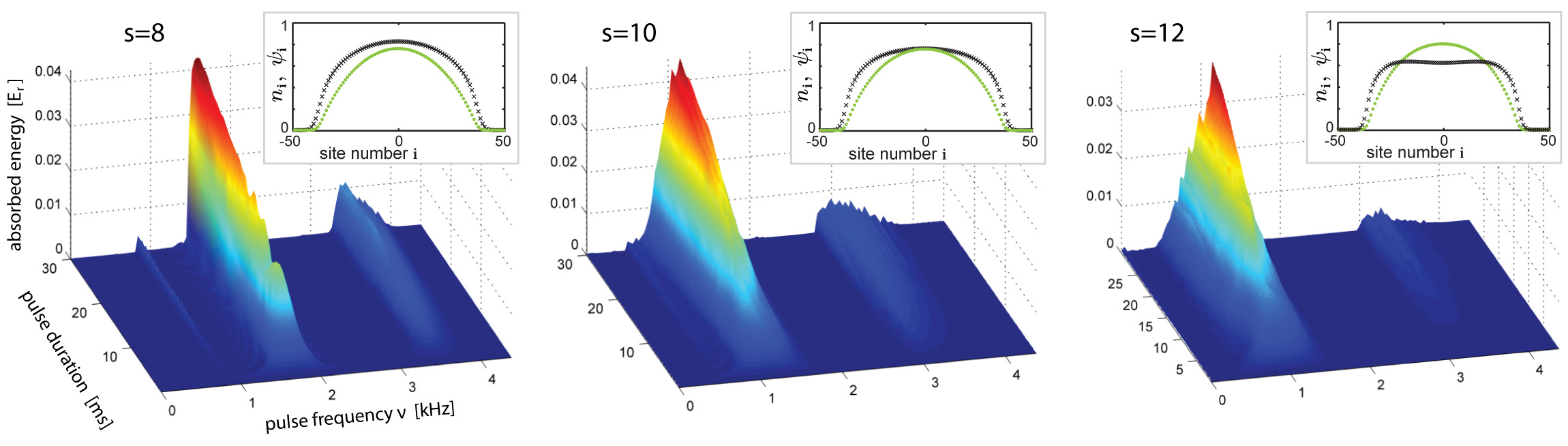}
\vspace{-7mm}
\end{figure}
FIG.~2. Absorbed energy in the condensate during lattice amplitude modulation for different lattice amplitudes $s=8,10,12$. Note that the resonance frequency of the lowest mode (amplitude mode at $k=0$) decreases with increasing lattice intensity (i.e. increasing interaction strength) in accordance with the experimental observation [5]. The insets show the ground state density profile $n_\bi$ (green dots) and the corresponding order parameter distribution $\psi_\bi$ (black crosses) along the x-direction through the center of the 3D trap, determined for the respective experimental parameters.
\end{widetext}

The absorbed energy, calculated within a time-dependent bosonic Gutzwiller calculation, is shown in Fig.~2 as a function of the modulation frequency and the exposure time. The parameters for this simulation were chosen in accordance with the experiment [5], i.e. $^{87}$Rb atoms in a $826$nm optical lattice were exposed to a modulation of $20\%$ of the lattice depth and the maximum shown time of $30$ms corresponds to the experimental exposure time. We considered three lattice intensities $s=8, 10, 12$ corresponding to different ratios of $U/J$, all on the condensate side of the superfluid-Mott insulator transition.  The initial density distribution was chosen, such that the central density agreed with the central density calculated for the actual experimental situation (insets in Fig.~2), with a total particle number of $N=1.5 \times 10^5$  in a lattice with an underlying $\omega=2\pi\cdot(18,20,22)\,$Hz harmonic trap and translational invariance was assumed in the $z$ direction. The frequencies of the absorption peaks in our simulation agree well with the position of the observed maxima for the different lattice intensities. The experimentally observed peak widths are, however, considerably larger than in our simulations, an effect which may be attributed to a finite temperature in experiment, as opposed to $T=0$ in the simulation. Moreover, an increase in the harmonic trapping frequency due to the lattice beams, which was not specified in [5] and accounted for in our simulation, may lead to a higher particle density at the center and hence also a greater inhomogeneity in the density profile, which would lead to a broadening of the absorption peaks. Due to this discrepancy in the peak width, we cannot claim to obtain real quantitative agreement with the experimental data, as for our Bragg experiment. The good correspondence in the peak positions does however indicate, that the lowest peak in the 3D measurement [5] can be associated with the amplitude mode at $k=0$.

We found that with lattice amplitude modulation spectroscopy the Bogoliubov sound mode cannot be excited, also when going beyond the linear response regime. Thus all peaks observed in the strongly interacting condensate can be attributed to the amplitude and higher gapped modes at $k=0$ and we furthermore point out that the larger width of the peaks cannot directly be attributed to the condensate phase. The peak ranging from $\nu=0.9$kHz to $1.9$kHz for different lattice intensities in our simulation can be identified with the amplitude mode at $k=0$, but seems to overlap and merges with the next mode in the experimental data. When crossing the phase transition into the Mott insulator, combined particle-hole modes appear at an energetically similar position as the amplitude mode. However, the quasi-particle structure fundamentally changes when crossing the superfluid-Mott insulator transition and we emphasize that the amplitude mode cannot simply be seen as the $U$-mode in the condensate. Whereas the phase response of a coherently excited amplitude mode state vanishes when approaching the transition from the superfluid side, this is large everywhere in the Mott insulator for both the particle and hole branches.
\end{document}